# Optical imaging with only the evanescent fields and without diffraction

Mufei Xiao

*Centro de Nanociencias y Nanotecnología, Universidad Nacional Autónoma de México,*

*km. 107 Carretera Tijuana-Ensenada, Ensenada, Baja California CP 22860, México*

*mufei@ens.cnyn.unam.mx*

**Abstract:** We report on a conceptual framework for optical imaging utilizing exclusively evanescent fields, effectively eliminating or attenuating standard optical diffraction. We present the theoretical foundation of this concept based on the decomposition of dipole radiation into evanescent fields and homogeneous propagating waves, exploring the explicit conditions required to suppress the latter. By successfully decoupling and removing the homogeneous component, optical tasks—such as sub-diffraction imaging—can be executed solely via evanescent channels, bypassing traditional diffractive limits. We discuss the physical feasibility of this approach and provide concrete examples of how this conjecture can be realized using currently available laboratory and industrial techniques. The realization of this concept promises to redefine core paradigms in optics and fundamental physics. Because the minimum achievable feature size is no longer governed by the operating wavelength, this approach enables an unprecedented increase in spatial resolution, offering a direct pathway to advance state-of-the-art semiconductor lithography and chip manufacturing.

## 1. Introduction

Light is fundamentally composed of electromagnetic waves and is therefore intrinsically subject to optical diffraction. This wave nature imposes a fundamental diffraction limit on the minimum achievable spatial feature size (light spot), which is strictly scaled by the operating wavelength $\lambda$.

An alternative paradigm to bypass the diffraction limit is near-field optics.[1] The electromagnetic field radiated by a localized source can be partitioned into two distinct components:

evanescent fields and homogeneous (propagating) waves. Evanescent fields decay exponentially away from the source and do not propagate as free waves; consequently, they are not subject to standard optical diffraction. However, these evanescent components are confined strictly to the near- and intermediate-field zones, which, for visible light, span only a few nanometers from the scatterer. Crucially, in conventional near-field optics, the propagating homogeneous waves coexist alongside the evanescent fields, meaning that diffraction effects are altered but not fundamentally eliminated. A long-standing objective in optical physics is to execute imaging and wave manipulation using *exclusively* the evanescent fields by selectively attenuating or completely suppressing the homogeneous wave spectrum. Achieving such a state would decouple spatial resolution from the operating wavelength, effectively rendering the system diffraction-free.

In this work, we propose a conceptual framework to realize this objective. The theoretical foundation relies on the exact angular spectrum decomposition of dipole radiation into its constituent evanescent and homogeneous wave components via a two-dimensional plane-wave expansion—specifically, a spatial Fourier transform projected onto a chosen reference plane.[2-5] While the initial reporting of this exact decomposition over two decades ago generated significant debate regarding its physical interpretation,[6] subsequent theoretical developments and experimental validations have firmly established its validity. Crucially, this decomposition is mathematically exact, requiring no approximations or reliance on truncated special functions.

Here, we exploit this exact formalism to investigate the suppression of homogeneous waves within localized spatial zones, focusing on a configuration featuring a primary dipole and its corresponding mirror image. Under tailored conditions, this dipole-mirror pair operates as a twin radiator system. By optimizing the phase and geometric spatial separation, the homogeneous propagating plane waves stemming from both dipoles can undergo total destructive interference, mirroring classical coherent wave superposition. In the regions where these homogeneous waves cancel out, classical diffraction phenomena vanish. This destructive suppression is localized within the near- and intermediate-field zones where the evanescent fields retain significant amplitude and the homogeneous components maintain strict phase coherence.

We analyze the experimental and industrial feasibility of this twin-dipole configuration. Given the precision of modern nanotechnology, such a system can be engineered in state-of-the-art laboratories using established fabrication and positioning tools, without requiring exotic materials or unachievable experimental conditions. While an ideal system yields total diffraction elimination, practical laboratory implementations will manifest as a highly controlled attenuation and localized reduction of the propagating field. The successful realization of this framework holds transformative potential for semiconductor lithography, offering chipmakers a direct pathway beyond EUV limits, while fundamentally opening a new epoch in nano-optics and radiation physics.

The remainder of this paper is organized as follows. In Section 2, the two-dimensional plane-wave expansion is formulated, providing a detailed, exact decomposition of dipole radiation. Section 3 analyzes the twin-radiator system composed of the primary and mirror dipoles, detailing the mechanisms for destructive homogeneous wave suppression. Section 4 evaluates alternative geometric designs extending beyond the ideal mirror-dipole model, demonstrating the robust feasibility of this framework under realistic laboratory conditions.

## 2. Decomposition of dipole radiation into evanescent fields and homogeneous waves

The decomposition is based on a two-dimensional plane-wave expansion.[2-5] This framework utilizes a spatial Fourier transform based on the projection of the three-dimensional spatial wavevector onto a selected reference plane $\vec{q}_{//}$.

Let us start with the radiation field at $\vec{r}$ in free space of a point dipole at the origin of dipole moment $\vec{P}$, as

$$\vec{E}(\vec{r}) = -\mu_0 \omega^2 \overleftrightarrow{G}(\vec{r}) \cdot \vec{p} \tag{1}$$

where $\mu_0$ is the vacuum permeability, $\omega$ the angular frequency of light, and $\vec{G}(\vec{r})$ is the dyadic field propagator tensor (Green's function) to describe the field distribution in all space. In angular frequency representation, the field propagator for a two-dimensional wavevector $\vec{q}_{//}$ can be written as:

$$\overleftrightarrow{\mathrm{G}}(\mathrm{z},\vec{\mathrm{q}}_{//}) = \frac{\exp(iq_z z)}{2iq_z}\left[\vec{\mathrm{e}}_{\mathrm{y}}\vec{e}_y + \vec{e}\vec{e}\right] \tag{2}$$

where vectors $\vec{e}_y = (0,1,0)$, $\vec{e} = (c/\omega)(-q_z,1,-q_{//})$, and $q_z^2 = (\omega/c)^2 - q_{//}^2$. The real space propagator is just the inverse Fourier transform of the plane wave in Eq. (2) for $0 \sim q_{//} \sim \infty$ as:

$$\overleftrightarrow{\mathrm{G}}(\vec{\mathrm{r}}) = \frac{1}{(2\pi)^2}\int_0^\infty \overleftrightarrow{S}^{-1} \cdot \overleftrightarrow{G}(z,\vec{q}_{//}) \cdot \overleftrightarrow{S} \exp(i\vec{q}_{//} \cdot \vec{r}_{//}) d^2 q_{//} \tag{3}$$

where $\vec{r}_{//}$ is in $\vec{q}_{//}$ plane, and $\overleftrightarrow{S}$ is a tensor to rotate the coordinate as:

$$\overleftrightarrow{S} = \frac{1}{q_{//}}\begin{bmatrix} q_x & q_y & 0 \\ -q_y & q_x & 0 \\ 0 & 0 & q_{//} \end{bmatrix} \tag{4}$$

The integration in Eq.(3) may be carried out in a simple process as follows. One assumes that $z$ is along the $\vec{r}$ direction. In this case the propagator along the $z$ - axis can be simply carried out because $\vec{r}_{//} = 0$ as

$$\overleftrightarrow{\mathrm{G}}(\mathrm{z}) = \frac{1}{4\pi}\left[\left(-\frac{1}{z} - \frac{ic}{\omega z^2} + \frac{c^2}{\omega^2 z^3}\right)\overleftrightarrow{U} + \left(\frac{1}{z} + \frac{3ic}{\omega z^2} - \frac{3c^2}{\omega^2 z^3}\right)\vec{e}_z\vec{e}_z\right]\exp\left(i\frac{\omega}{c}z\right) \tag{5}$$

where $\overleftrightarrow{U}$ is a unit tensor and $\vec{e}_z$ is the unit vector along the $z$ - coordinate. Inserting Eq.(5) into Eq.(1), one realizes that the first term in Eq.(5) shows a field along the dipole direction, and the second term in Eq.(5) is a field with a direction determined by the direction of the dipole, as well as the observation direction. Since there is no restriction for the selection of the dipole and observation directions, one is able to rewrite the expression in $z = r$ and $\vec{e}_z = \vec{n}$ ($\vec{n}$ - unit vector in $\vec{r}$ direction), and the expression for all space becomes

$$\vec{\vec{\mathrm{G}}}(\vec{\mathrm{r}}) = \frac{1}{4\pi}\left[\left(-\frac{1}{r} - \frac{ic}{\omega r^2} + \frac{c^2}{\omega^2 r^3}\right)\vec{\vec{U}} + \left(\frac{1}{r} + \frac{3ic}{\omega r^2} - \frac{3c^2}{\omega^2 r^3}\right)\vec{n}\vec{n}\right]\exp\left(i\frac{\omega}{c}r\right) \tag{6}$$

Inserting Eq.(6) into Eq.(1), one retrieves the well-known dipole field in textbooks. This confirms that we have carried out a correct and rigorous transfer from a two-dimensional angular representation to a real space field distribution. It is now safe to employ the same mathematical process to obtain the expression for an evanescent field propagator just by changing the integration range to $\omega / c \sim q_{//} \sim \infty$. In this range, the plane waves in Eq.(2) are evanescent as they decay exponentially. One integrates straightforwardly with $\vec{r}_{//} = 0$ and then rewrites the expression with $z = r$ and $\vec{e}_z = \vec{n}$ to obtain the expression for all space. The resulting evanescent field propagator can be written as:

$$\vec{\vec{\mathrm{G}}}_{\mathrm{evan}}(\vec{\mathrm{r}}) = \frac{1}{4\pi}\left[\left(-\frac{1}{2r} + \frac{c^2}{\omega^2 r^3}\right)\vec{\vec{U}} + \left(-\frac{1}{2r} - \frac{3c^2}{\omega^2 r^3}\right)\vec{n}\vec{n}\right] \tag{7}$$

and the homogenous field propagator is

$$\vec{\vec{\mathrm{G}}}_{\mathrm{homo}}(\vec{\mathrm{r}}) = \vec{\vec{\mathrm{G}}}(\vec{\mathrm{r}}) - \vec{\vec{\mathrm{G}}}_{\mathrm{evan}}(\vec{\mathrm{r}}) \tag{8}$$

In the subsequent section, a conceptual framework will be presented based on the decomposition revisited in the present section. Below, we take three notes about the decomposition. The notes are essential and crucial for understanding the decomposition and the concept of diffraction free to be proposed in the subsequent section.

### 2.1 The singularity is borne entirely by the evanescent fields

A singularity at the origin $r \to 0$ is intrinsic to a point source model. One expects that the evanescent component carries the entire singularity while the homogenous waves are finite at $r \to 0$. This serves as a crucial consistency test to verify the physical validity of the spatial spectrum decomposition.. One takes a Maclaurin expansion of $\exp\left(i\frac{\omega}{c}r\right)_{r\to 0}$ and carefully evaluates the two components $\overleftrightarrow{G}_{evan}(r\to 0)$ and $\overleftrightarrow{G}_{\mathrm{homo}}(r\to 0)$. The evanescent field propagator eventually shows a singularity $\overleftrightarrow{G}_{evan}(r\to 0)=\infty$, whereas the homogeneous field propagator becomes finite

$$\overleftrightarrow{G}_{\mathrm{homo}}(r\to 0)=\frac{-i\omega}{6\pi c}\overleftrightarrow{U}=\frac{-i}{3\lambda}\overleftrightarrow{U} \tag{9}$$

Note also that the above expression in Eq.(9) is an invariant function of the operating wavelength only. The tests on the total field propagator, the homogeneous field propagator, and the evanescent field propagator when $r\to 0$ confirms the correctness and rigorousness of the decomposition.

**2.2 The decomposition is mathematically exact and rigorous**

It is worth emphasizing that this mathematical partition is exact and free of any approximations whatsoever. For alternative tasks in near-field optics, two-dimensional plane-wave expansions often rely on asymptotic approximations or paraxial truncations to evaluate paraxial or non-paraxial fields. In our case, no such mathematical assumptions are required, as we consistently account for the entire spatial frequency spectrum. For variations in the observation trajectory, the reference plane can be dynamically adjusted to repeat the identical Fourier projection process. Consequently, the results presented here are valid for all arbitrary directions and encompass all spatial coordinates, establishing a rigorous framework devoid of empirical scaling factors.

**2.3 The evanescent fields include all high spatial frequencies**

This note is foundational to understanding the concept of diffraction-free imaging introduced in the subsequent section. A common misconception in near-field optics is to treat the evanescent field as physically equivalent to a quasi-static electrostatic field. While electrostatic approximations occasionally yield acceptable practical values in specialized regimes,[7,8] this assumption is fundamentally incomplete. The decomposition verified in this work establishes that the evanescent component encapsulates all high spatial frequencies ($\omega/c\sim q_{//}\sim\infty$). Conversely, the homogeneous component encompasses low spatial frequencies ( $0 \sim q_{//} \sim \omega / c$ ), which uniquely share the radiating spectrum and include the far-field propagation vectors. Correctly recognizing this distinction clarifies how manipulating the propagating spectrum can successfully decouple near-field resolution from the free-space wavelength limits.

### 3. Ananalytical framework to remove diffraction

In the previous section, we established the exact mathematical framework to decompose dipole radiation into its constituent evanescent and homogeneous components. Building upon this decomposition, we now propose a conceptual framework to selectively attenuate or completely eliminate the homogeneous wave vector component, thereby isolating pure evanescent channels for optical imaging. Because pure evanescent fields are entirely unconstrained by far-field wave propagation constraints, standard optical diffraction limits vanish, removing the fundamental wavelength barrier to spatial resolution.

To date, the literature contains no validated mechanisms for completely diffraction-free optical imaging. The historical consensus holds that evanescent fields and homogeneous waves are inextricably linked in open environments. In classical near-field scanning optical microscopy (NSOM), sub-wavelength resolutions are achieved simply because the probe isolates the high amplitudes of evanescent fields within a few nanometers of the sample. However, homogeneous background waves always coexist, generating unavoidable diffractive scattering. While alternative approaches have attempted to mathematically subtract diffraction signatures by processing time-resolved sequential images,[9] these post-processing routines merely attenuate background noise relatively rather than eliminating propagating diffraction modes at the physical onset.

In this work, we propose removing the propagating wave component at the structural onset. We begin with an analytical model assuming a paired twin-radiator configuration at the position of the localized emitter. If two localized dipoles radiate coherently with an absolute phase shift of π (opposite phases), their matching low-frequency spatial plane waves can undergo total destructive interference, mirroring classical coherent far-field wave superposition. Analyzing this interaction via a two-dimensional plane-wave expansion reveals that the propagating modes can be completely canceled out, leaving behind an isolated near-field zone dominated exclusively by high-frequency evanescent fields. At least along the designated projection axes, the localized interaction volume becomes fundamentally diffraction-free.

We propose a practical physical realization of this architecture utilizing a primary dipole-mirror configuration. When a primary dipole radiator is placed at a deep sub-wavelength distance from an engineered reflective interface, the boundary reflection behaves precisely as a coherent, phase-shifted mirror dipole image. This primary-mirror pairing successfully fulfills the operational constraints of a coupled twin-radiator system. Within this configuration, several geometric and material parameters can be tuned to maximize field suppression: the dipole-to-mirror separation distance, the localized polarizability (strength) of the primary emitter, and the complex reflectivity spectrum of the boundary interface. Furthermore, if the spatial coupling between the dipole near-field and the surface electronic states is sufficiently strong, the system can enter a resonant radiation regime. To accurately model these strong localized coupling states, a self-consistent field solution must be employed.[10,11,12]

Tailoring this mirror-dipole environment is entirely feasible within modern nanotechnology and nanofabrication laboratories, presenting an actionable roadmap for industrial processes such as chip manufacturing. While an absolute mathematical zero in propagation represents an idealized limit, realistic laboratory implementations can achieve a highly controlled, profound attenuation of the propagating background. A recognized constraint of this basic planar setup is that the total elimination of homogeneous wave vectors occurs preferentially along the vertical alignment axis, rendering a narrow pure-evanescent field window. In Fig.1, the idea is schematically explained. To broaden this spatial working window, we will outline a series of alternative, optimized geometric configurations in the following section.

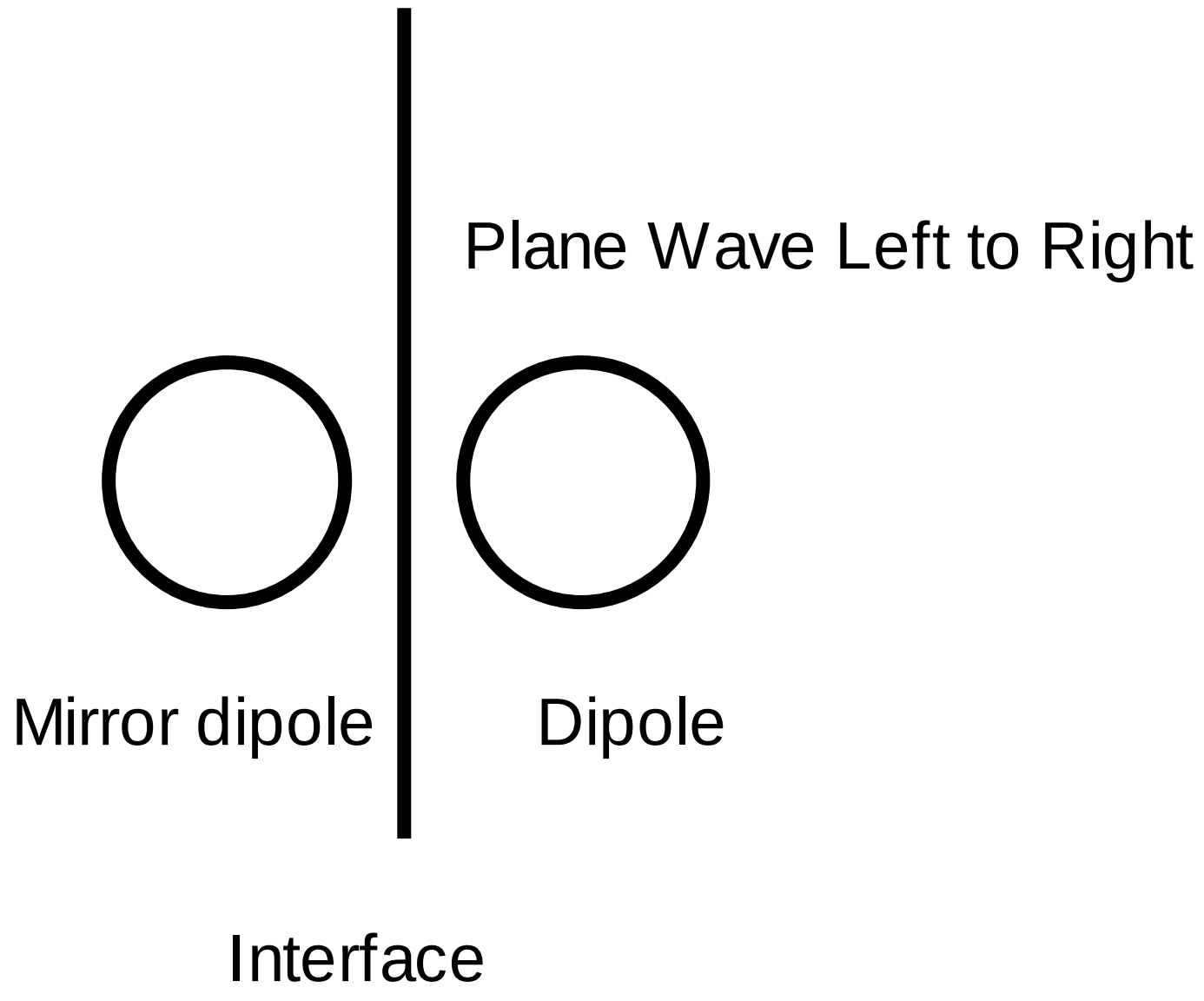

(a)

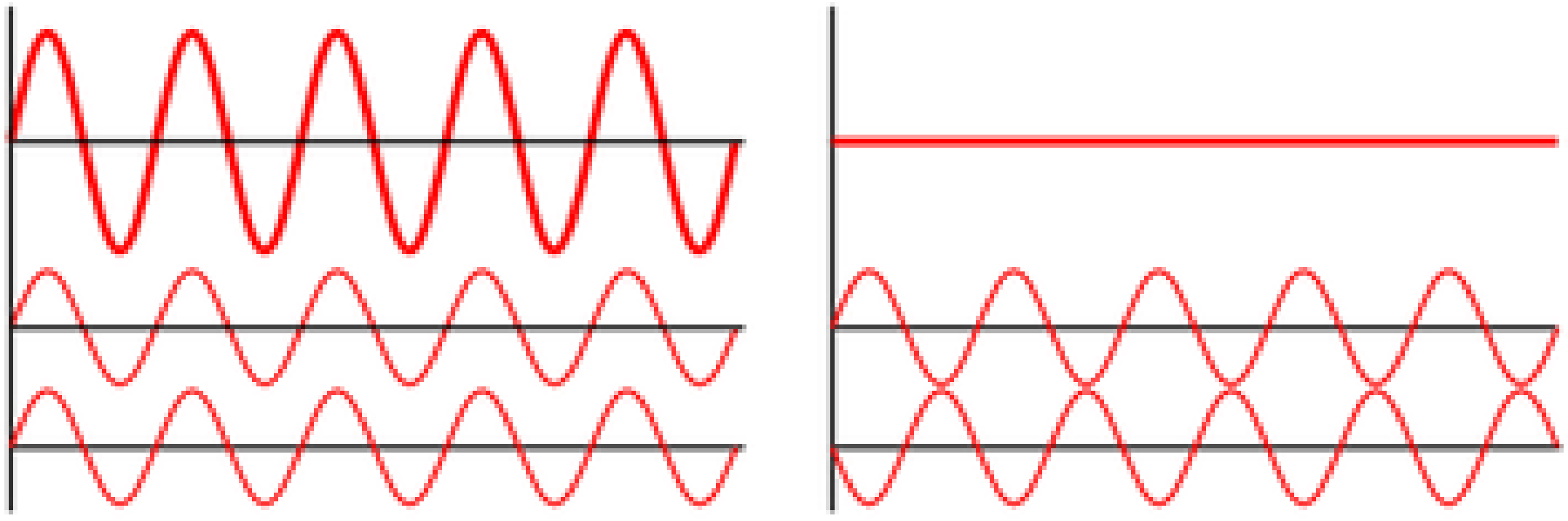

(b)

Fig. 1.Schematic layout of the coupled twin-radiator framework. (a) A primary localized dipole coupled to an engineered interface with reflectivity R, establishing a coherent, phase-shifted mirror image dipole. Key parameters include separation distance, source polarizability, and interface reflectivity. (b) Representation of plane-wave spectrum interference: constructive conditions (left) versus destructive cancellation conditions (right), illustrating the systematic elimination of the homogeneous propagating spectrum.

## 4. Alternative configurations and future perspectives

In this work, we have outlined a comprehensive conceptual framework for optical imaging utilizing exclusively evanescent fields. By successfully decoupling and suppressing the low-spatial-frequency homogeneous wave spectrum, standard far-field optical diffraction effects can be effectively eliminated or significantly attenuated at the structural source. The exact theoretical foundation relies on partitioning dipole radiation via a two-dimensional angular spectrum expansion, establishing explicit boundary conditions necessary to cancel out propagating modes. Given the precision of modern nanotechnology, this conjecture can be realized using available laboratory and industrial techniques, such as the localized twin-radiator configuration. Realizing this concept promises to redefine core paradigms in nano-optics and fundamental radiation physics. Because the minimum achievable feature size is decoupled from the operating wavelength, this approach enables an unprecedented increase in spatial resolution, offering a direct pathway to advance state-of-the-art semiconductor lithography and chip manufacturing.

While precise experimental validation from specialized laboratories remains the critical next phase, we outline below several alternative geometric configurations beyond the basic planar twin-radiator model. These alternative designs are intended to expand the operational parameter space, maximize the purely evanescent working window, and provide flexible implementation paths focusing on source geometry and boundary conditions.

### 4.1 Emitter geometry and spatial detection profiles

Although the exact spatial frequency decomposition detailed in Section 2 is formulated using a localized point-dipole source model, practical physical radiators do not need to be strictly spherical or point-like. Utilizing an extended, asymmetric radiator morphology aligned along the boundary interface can significantly widen the spatial working window. For instance, the source can be engineered as an elongated nanoscale needle or an extended conductive wire. Systematic near-field scanning localized directly above these optimized geometries can map and isolate the target pure-evanescent interaction zones. Furthermore, a sharp needle-tip geometry can serve as a localized hot-spot for enhanced evanescent field collection, while an integrated network or periodic array of coupled radiators can be engineered to systematically enhance the signal-to-noise ratio of the sub-wavelength channels. These alternative layouts can be readily analyzed via numerical simulations and tailored to specific experimental constraints using existing lithographic tools.

### 4.2 Core cavity structures and engineering mirror materials

The reflective boundary or mirror component can also be optimized using non-planar geometries. The primary dipole radiator can be structurally buried within a stratified medium or embedded directly inside an engineered micro-cavity. Enclosing a localized dipole source within a highly reflective spherical shell, for example, dramatically increases the geometric paths available to enforce destructive interference of propagating modes. In such a setup, a deep sub-wavelength aperture or slot cut into the shell can act as an isolated near-field outlet, routing the pure evanescent signals outward while trapping propagating modes inside. The core objective of these advanced structural designs is to maximize the localized near-field interaction area while minimize the angular escape window of the diffractive propagating field.

Furthermore, the boundary materials can be engineered with specialized spectral selectivity or custom dielectric tensor layouts to ensure maximum reflectivity precisely at the operating frequency of the source. Crucially, initial experimental verification of this framework does not depend on immediate sub-micron nanofabrication. Proof-of-concept validations can be efficiently executed at macroscopic scales using radio frequencies (RF) or microwave regimes, where complex macro-structures can be easily fabricated and measured with standard laboratory equipment.

### 4.3 On-chip interconnects and guided optical coupling

While the primary industrial impact of diffraction-free imaging is naturally situated within deep sub-wavelength photolithography, this framework addresses another critical frontier in semiconductor engineering: localized optical coupling. Light is increasingly utilized within high-performance microchips for internal data communications, systematically replacing traditional copper wires to bypass ohmic losses. High-efficiency coupling between adjacent optoelectronic devices is essential to relay data channels across a dense chip architecture. If this localized light transportation can be restricted entirely to pure evanescent channels, far-field radiative leakage and crosstalk can be prevented at the structural onset. Consequently, next-generation optoelectronic integrated circuits (ICs) can be designed with tighter component densities while completely eliminating parasitic signal leakage.